\documentclass[conference]{IEEEtran}
\IEEEoverridecommandlockouts

\usepackage[a4paper, left=1.6cm, right=1.6cm, top=1.8cm, bottom=4.4cm, columnsep=0.21in]{geometry}

\usepackage{cite}
\usepackage{amsmath,amssymb,amsfonts}
\usepackage{graphicx}
\usepackage{epstopdf}
\usepackage{algorithm}
\usepackage{algorithmic}
\usepackage{subfigure}
\usepackage{textcomp}
\usepackage{xcolor}
\usepackage{float}
\usepackage{balance}
\usepackage{setspace}
\def\BibTeX{{\rm B\kern-.05em{\sc i\kern-.025em b}\kern-.08em
    T\kern-.1667em\lower.7ex\hbox{E}\kern-.125emX}}
\begin{document}

\title{Symbol Timing Synchronization and Signal Detection for Ambient Backscatter Communication \\
}

\author{
    \IEEEauthorblockN{Yuxin Li\IEEEauthorrefmark{1}, Guangyue Lu\IEEEauthorrefmark{1}, Yinghui Ye\IEEEauthorrefmark{1}, Zehui Xiong\IEEEauthorrefmark{2}, and Liqin Shi\IEEEauthorrefmark{1}}
    \IEEEauthorblockA{\IEEEauthorrefmark{1}Shaanxi information communication network and security Laboratory,\\ Xi'an University of Posts and Telecommunications, Xi'an, China\\\IEEEauthorrefmark{2}School of Electronics, Electrical Engineering and Computer Science, Queen's University Belfast, Belfast, United Kingdom\\
    Email: liyuxin@stu.xupt.edu.cn, tonylugy@163.com, connectyyh@126.com,\\ z.xiong@qub.ac.uk, liqinshi@hotmail.com}
    \thanks{This work was supported in part by the National Natural Science Foundation of China under Grant 62471388, in part by the Innovation Capability Support Program of Shaanxi under Grant 2024ZC-KJXX-016, and in part by the Key Research and Development Program of Shaanxi under Grant 2024GX-ZDCYL-01-32.}
}\maketitle

\begin{abstract}
Ambient backscatter communication (AmBC) enables ambient Internet of Things (AIoT) devices to achieve ultra-low-power, low-cost, and massive connectivity. Most existing AmBC studies assume ideal synchronization between the backscatter device (BD) and the backscatter receiver (BR). However, in practice, symbol timing offset (STO) occurs due to both the propagation delay and the BR activation latency, which leads to unreliable symbol recovery at the BR. Moreover, the uncontrollable nature of the ambient radio frequency source renders conventional correlation-based synchronization methods infeasible in AmBC. To address this challenge, we investigate STO estimation and symbol detection in AmBC without requiring coordination from the ambient radio frequency source. Firstly, we design a specialized pilot sequence at the BD to induce sampling errors in the pilot signal. Furthermore, we propose a pilot-based STO estimator using the framework of maximum likelihood estimation (MLE), which can exploit the statistical variations in the received pilot signal. Finally, we integrate STO compensation into an energy detector and evaluate the bit error rate (BER) performance. Simulation results show that the proposed estimator achieves accurate STO estimation and effectively mitigates the BER performance degradation caused by STO.
\end{abstract}


\section{Introduction}
The accelerating growth of the Internet of Things (IoT) is reshaping modern life and industry, while creating a pressing demand for communication technologies that offer low power consumption and reduced deployment costs. Ambient backscatter communication (AmBC) has emerged as a key enabler for future sustainable ambient IoT (AIoT) deployments\cite{9509294,10463656,8368232}. Unlike conventional communications that rely on dedicated carrier emitters, AmBC allows the backscatter device (BD) to leverage ambient radio frequency (RF) signals from legacy systems to communicate with the backscatter receiver (BR). Specifically, the BD harvests energy from the ambient RF sources to operate and transmit information by dynamically switching its antenna impedance between absorbing and reflecting states\cite{9812481,9055221}. By eliminating the need for active RF transmission and external power supply, AmBC significantly reduces the deployment and maintenance costs of AIoT. However, in this communication paradigm,  the BD's weak backscatter signal suffers from a double-path loss and is easily overwhelmed by strong ambient RF signals, which makes it challenging for symbol detection at the BR\cite{10915625}.

Most existing works address this issue under the assumption of ideal synchronization at the BR\cite{liu2013ambient,7341107,8007328,wang2016ambient}. In \cite{liu2013ambient}, AmBC was first introduced with a hardware prototype, where BD symbol detection relied on the difference in modulation rates between the incident RF signal and the BD signal. Based on this pioneering work, several theoretical studies were conducted, which can be broadly categorized into three types: coherent detection, semi-coherent detection, and non-coherent detection. Coherent detection allows the BR to achieve high sensitivity in detecting weak BD signals, but it requires accurate channel state information (CSI). For example, the authors of \cite{7341107} proposed a coherent maximum likelihood (ML) detector and derived the optimal detection threshold and bit error rate (BER) in closed-form expressions. Semi-coherent detection avoids the dependence on CSI by utilizing a limited number of pilot symbols to estimate the necessary parameters for calculating the detection threshold, as explored in \cite{8007328}. To eliminate the need for prior information, non-coherent detection was investigated. Specifically, by leveraging differential coding schemes, the authors of \cite{wang2016ambient} proposed an energy difference detector and derived the corresponding near-optimal detection threshold and BER expressions. However, symbol timing offset (STO) inevitably occurs due to propagation delay and the BR activation latency, making it difficult to achieve ideal synchronization. As a result, the sampling window of the current received symbol includes parts of adjacent symbol, known as sampling mismatch, which can cause the sampling errors and significantly degrade the detection performance.

To this end, several studies have investigated synchronization for AmBC in the presence of STO\cite{8103807,10485514,10167801}. In \cite{8103807}, a correlation-based synchronization scheme was proposed, but its high computational complexity rendered it impractical for a resource-constrained BR. Later, two estimators based on the ML and expectation-maximization algorithms were proposed in \cite{10485514}. Despite their accuracy, the approaches in \cite{8103807} and \cite{10485514} required the ambient RF source to be controllable and able to transmit dedicated pilots for the AmBC system. To reduce the dependence on the dedicated pilot signals from the ambient RF source, the authors of \cite{10167801} leveraged the existing synchronization sequences and their periodicity in LTE signals. Nonetheless, this approach is only applicable to LTE-based systems. Overall, these methods are fundamentally incompatible with the inherent characteristics of AmBC, where the ambient RF source is typically unknown and operates without coordination. Since the symbols of the ambient RF source are unknown, the BD pilot signal modulated onto it is also unknown to the BR. As a result, conventional STO estimation methods based on sequence correlation are ineffective in AmBC\cite{852929,930629}.

Motivated by the aforementioned challenges, this paper investigates STO estimation and symbol detection without relying on either coordination with the ambient RF source. The main contributions of this paper are summarized as follows:
\begin{itemize}
  \item We design a novel pilot sequence that composed of alternating ``0'' and ``1'' bit-pairs to introduce sampling errors, thereby facilitating STO estimation.
  \item By leveraging the statistical variations in the received pilot signal induced by sampling errors, we develop an STO estimator based on the ML estimation (MLE), which does not rely on prior knowledge of the ambient RF source.
  \item We incorporate the estimated STO into an energy detection (ED) detector to perform compensation. Simulation results demonstrate the effectiveness of our proposed STO estimator, and BER degradation caused by STO can be significantly mitigated.
\end{itemize}


\section{System Model}
\begin{figure}
\centerline{\includegraphics{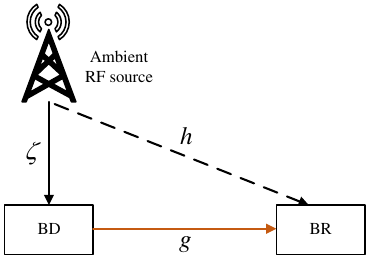}}
\caption{AmBC system.}
\label{fig1}
\end{figure}

Fig. \ref{fig1} depicts the AmBC network that consists of an ambient RF source (e.g., a cellular base station or a WiFi access point), a BD, and a BR, where the ambient RF source is completely unknown to and uncoordinated with both the BD and the BR.
In this network, the BD employs on-off keying (OOK) modulation at a lower rate than the ambient RF signal to backscatter its information onto the incident waveform, thereby transmitting bits ``1'' and ``0''. Let $h$, $\zeta$ and $g$ denote the channel fading coefficients between the ambient RF source and the BR, between the ambient RF source and the BD, and between the BD and the BR, which are assumed to be frequency-flat and block-fading, remaining constant within a coherence interval and varying independently across different intervals \cite{liu2013ambient,7341107,8007328,wang2016ambient}. Given the unknown nature of the ambient RF source, the signal $s(n)$ is modeled as an independent and identically distributed (i.i.d.) complex Gaussian process with zero mean and variance $\sigma_s^2$, i.e., $s(n) \sim \mathcal{CN}(0, \sigma_s^2)$. The thermal noise at the BR is typically modeled as a complex Gaussian signal, $ w(n) \sim \mathcal{CN}(0, \sigma_w^2)$. For the case of ideal synchronization (i.e., without STO), the received signal corresponding to the $k$-th symbol at the BR is given by\cite{8007328}
\begin{equation}
{y_k^{{\rm{ideal}}}}\left( n \right){\rm{ = }}\underbrace {hs\left( n \right)}_{{\rm{the}}\;{\rm{direct}}\;{\rm{link}}} + \underbrace {\zeta gB\left( k \right)s\left( n \right)}_{{\rm{the}}\;{\rm{backscatter}}\;{\rm{link}}} + \underbrace {w\left( n \right)}_{{\rm{noise}}}
\tag{1}
\label{eq1},
\end{equation}
where $n= kN+1,\dots,(k+1)N$, $N$ denotes the sample size for each symbol, and $B(k) \in \{0, 1\} $ represents the transmission of bit ``0'' or bit ``1'' by the BD. Accordingly,  the received signal can be modeled as a complex Gaussian random variable with the following distribution
\begin{equation}
{y_k^{{\rm{ideal}}}}(n) \sim \left\{ \begin{array}{l}
\mathcal{CN}(0,{P_0}),\ B(k) = 0\\
\mathcal{CN}(0,{P_1}),\ B(k) = 1
\end{array} \right.
\tag{2}
\label{eq2},
\end{equation}
where ${P_0} = {\left| h \right|^2}\sigma _s^2 + \sigma _w^2$, ${P_1} = {\left| \mu \right|^2}\sigma _s^2 + \sigma _w^2$, and $\mu = h+\zeta g$.
\subsection{The Impact of STO}\label{A}
In practical AmBC systems, STO inevitably occurs at both the BD and the BR, primarily due to propagation delays and the wake-up response latency\footnote{Energy-based rising edge detection is widely used for wake-up triggering in AmBC due to its simplicity; however, it inevitably incurs STO as a result of the early or delayed detection of the energy rising edge, with the offset typically bounded by the energy detection window width \cite{zhang2016hitchhike,dunna2021syncscatter}.} of the low-power devices. However, since the symbol duration of the BD signal is significantly longer than that of the ambient RF signal, the STO at the BD can be reasonably neglected. In the following, we mainly focus on the STO at the BR.
\begin{figure}
\centerline{\includegraphics[width=0.45\textwidth]{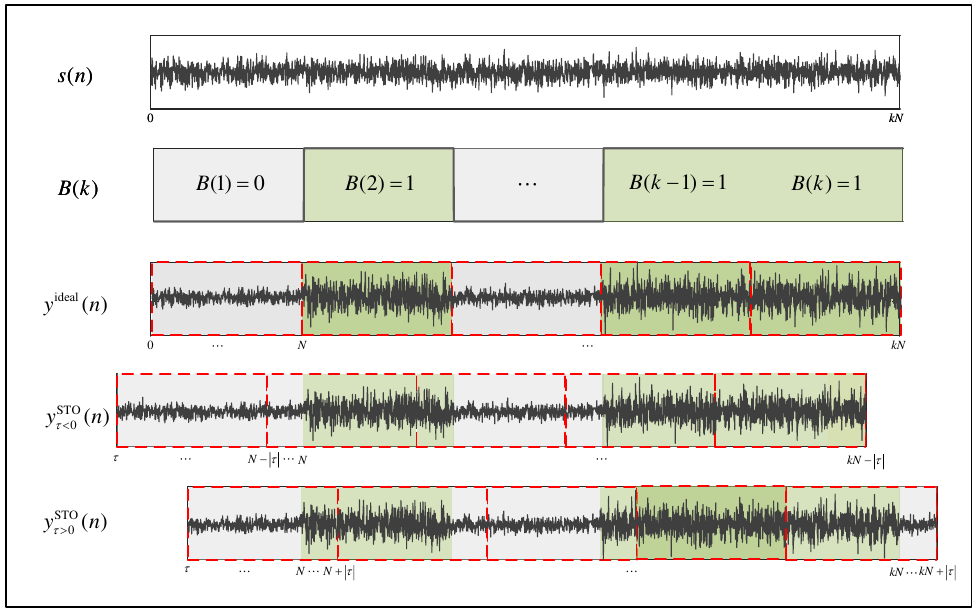}}
\caption{The received signal under ideal synchronization and non-ideal synchronization with STO.}
\label{fig2}
\end{figure}

In the presence of STO, the initial sampling time at the BR may be either advanced or delayed relative to the actual symbol arrival time, corresponding to $\tau<0$ and $\tau>0$, respectively. Fig. \ref{fig2} illustrates the impact of STO on the received signal, where $y_{\tau  < 0}^{{\rm{STO}}}\left( n \right)$ and $y_{\tau  > 0}^{{\rm{STO}}}\left( n \right)$ represent the received signal under sampling advance and sampling delay, respectively. As shown in Fig. \ref{fig2}, due to STO, the signals $y_{\tau < 0}^{{\rm{STO}}}\left( n \right)$ and $y_{\tau  > 0}^{{\rm{STO}}}\left( n \right)$ include samples from adjacent symbols within each sampling window. Specifically, for the case of $\tau <0$, the sampling window for the current symbol $B(k)$ includes part of the previous symbol $B(k-1)$. For the case of $\tau >0$, part of the following symbol $B(k+1)$ is included in the sampling window of the current symbol $B(k)$. If the adjacent symbol is the same bit value as the current one, i.e., $B(k)=B(k-1)$ for $\tau<0$ and $B(k)=B(k+1)$ for $\tau>0$, the received samples within the current sampling window will not be distorted by STO. However, when the adjacent symbol differs from the current one, i.e., $B(k) \ne B(k-1)$ for $\tau<0$ and $B(k) \ne B(k+1)$ for $\tau>0$, the sampling window contains samples from two symbols carrying opposite bit values, thereby introducing sampling errors and causing the received samples to no longer accurately reflect the characteristics of the current symbol. Notably, as $\left|\tau\right|$ approaches or exceeds half the sampling window, the received samples become dominated by the incorrect symbol, resulting in nearly complete detection error.

{\it{Remark 1:}} The above analysis reveals that sampling errors occur only when the current symbol differs from its adjacent symbols, which motivates the design of a pilot structure composed of alternating ``0'' and ``1'' bit-pairs to intentionally introduces symbol transitions and sampling errors. By examining the resulting variations in the received pilot signal, it becomes feasible to estimate the STO.
\section{BD Pilot Design, STO Estimation And Symbol Detection}
In this section, we present the architecture of the proposed BD pilot and illustrate how to achieve symbol timing synchronization between the BD and the BR.
\subsection{BD Pilot Architecture}
\begin{figure}
\centerline{\includegraphics[width=0.45\textwidth]{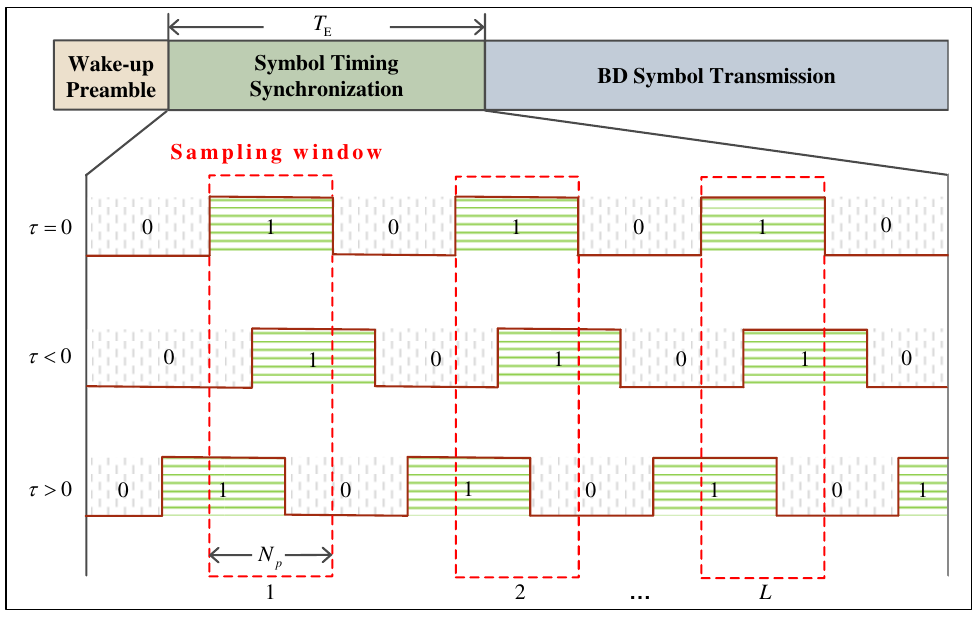}}
\caption{BD Pilot design for AmBC.}
\label{fig3}
\end{figure}

Due to the low-power and low-complexity requirements of AmBC, the pilot should be simple and avoid complex operations or computations. To this end, a practical pilot structure is proposed for the AmBC system, as illustrated in Fig. \ref{fig3}. Each transmission initiated by the BD consists of three sequential phases: a wake-up preamble phase, a symbol timing synchronization phase, and a BD data transmission phase.
\begin{itemize}
\item Wake-up Preamble Phase: In this phase, the BD continuously transmits an all-one sequence as a wake-up preamble to activate the corresponding BR. The BR employs rising edge detection\cite{zhang2016hitchhike,dunna2021syncscatter}, which identifies abrupt increases in the received signal energy to determine activation. As the rising edge detection relies on the accumulation of energy over a certain period, it inevitably introduces a response delay. As a result, the BR is unable to accurately identify the start of the BD's first data symbol, leading to STO.
\item Symbol Timing Synchronization Phase: The BD begins to transmit a specially designed sequence of alternating  ``0'' and ``1'' bits to the BR for STO estimation, where the total number of alternating bit-pairs is $L$ and each bit occupies $N_p$ samples. To ensure the detection of both $\tau<0$ and $\tau>0$, $N_p$ must satisfy\footnote {In practical implementations, the length of $N_p$ can be appropriately adjusted to guarantee compliance with this requirement, thereby keeping $N_p/2$ greater than $|\tau|$.} $N_p > 2\left |\tau\right|$. Meanwhile, the BR receives the pilot signal using a slide window of width $N_p$ and collects $L$ pilot signal segments at intervals of $N_p$ samples. After the pilot transmission, the received pilot signal can be represented as a matrix ${{\bf{Y}}_{L \times {N_p}}} = {\left[ {{{\bf{y}}_1},{{\bf{y}}_2}, \ldots ,{{\bf{y}}_L}} \right]^T}$, where ${{\bf{y}}_l} = \left[ {{y_l}(1),{y_l}(2), \ldots ,{y_l}({N_p})} \right]$ and $l = 1,2, \ldots ,L$.
\item BD Symbol Transmission Phase: During this phase, the BD transmits its data symbols to the BR  until all the data symbols have been transmitted.
\end{itemize}

It is important to ensure that the symbol timing synchronization duration, denoted as $T_{\rm{E}}$, should be within the coherence time of the channel, during which the channel coefficients are assumed to be constant.
\subsection{STO Estimation Using MLE}
Under ideal synchronization without STO, the received pilot signal within the sampling window corresponds entirely to bit ``1''. However, under STO, the sampling window at the BR spans a symbol transition, resulting in the inclusion of some samples corresponding to bit ``0''. Specifically, for the case of $\tau <0$, the received pilot signal in the $l$-th sampling window during can be given by
\begin{equation}
y_l^{\tau  < 0}(n) = \left\{ \begin{array}{l}
hx_l(n) + w_l(n),\ n = 1, \ldots ,\left| \tau  \right|\\
\mu x_l(n) + w_l(n),\ n = \left| \tau  \right| + 1, \ldots ,{N_p}
\end{array} \right.
\tag{3}
\label{eq3},
\end{equation}
where $n = 1,2,\ldots,N_p$, $x_l(n)$ and $w_l(n)$ represent the ambient RF source signal and noise signal of the $l$-th sampling window, respectively.

For the case of $\tau >0$, the received pilot signal in the $l$-th sampling window can be given by
\begin{equation}
y_l^{\tau  >0}(n) = \left\{ \begin{array}{l}
\mu x_l(n) + w_l(n),\ n = 1, \ldots ,N_p - \left| \tau  \right|\\
h x_l(n) + w_l(n),\ n = N_p - \left| \tau  \right| + 1, \ldots ,{N_p}
\end{array} \right.
\tag{4}
\label{eq4}.
\end{equation}
\begin{figure}
\centerline{\includegraphics[width=0.38\textwidth]{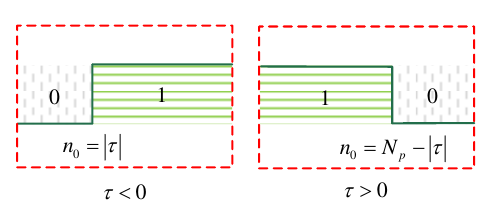}}
\caption{The received pilot signal under STO within a sampling window.}
\label{fig4}
\end{figure}

To facilitate analysis, Fig. \ref{fig4} illustrates the received pilot signal in the $l$-th sampling window for the cases of $\tau < 0 $ and $\tau >0$, respectively, where the signal transition point corresponding to the STO is denoted by $n_0$. Specifically, $n_0 = \left| \tau  \right|$ when $\tau < 0$, and $n_0 = N_p-\left| \tau  \right|$ when $\tau > 0$. Since the BR is unable to determine whether the sampling is advanced or delayed, we divide the sampling window into two segments. The CSI over the interval $\left[ {1,{n_0}} \right]$ is denoted as $\alpha_1$, and over the interval $\left[ {{n_0+1},{N_p}} \right]$ as $\alpha_2$. Both $\alpha_1$ and $\alpha_2$ are unknown to the BR. Based on this, the signal models in (\ref{eq3}) and (\ref{eq4}) can be unified and simplified as follows
\begin{equation}
y_l^{\rm{STO}}(n) = \left\{ \begin{array}{l}
\alpha_{1} x_l(n) + w_l(n),\ n = 1, \ldots ,n_0\\
\alpha_{2} x_l(n) + w_l(n),\ n = n_0 + 1, \ldots ,{N_p}
\end{array} \right.
\tag{5}
\label{eq5},
\end{equation}
with the distribution given by
\begin{equation}
{y_l^{{\rm{STO}}}}(n) \sim \left\{ \begin{array}{l}
\mathcal{CN}(0,{\sigma_1^2}),\ n = 1, \ldots ,n_0\\
\mathcal{CN}(0,{\sigma_2^2}),\ n = n_0 + 1, \ldots ,{N_p}
\end{array} \right.
\tag{6}
\label{eq6},
\end{equation}
where ${\sigma_1^2}$ and ${\sigma_2^2}$ are the variances of ${y_l^{{\rm{STO}}}}(n)$ over the intervals $\left[ {1,{n_0}} \right]$ and $\left[ {{n_0+1},{N_p}} \right]$, respectively. Specifically, ${\sigma_1^2} = {\left| \alpha_1 \right|^2}\sigma _s^2 + \sigma _w^2$, and ${\sigma_2^2} = {\left| \alpha_2 \right|^2}\sigma _s^2 + \sigma _w^2$.

In addition to $\alpha_1$ and $\alpha_2$, the powers of the ambient RF source signal and noise, $\sigma_s^2$ and $\sigma_w^2$,  are also unknown to the BR. Consequently, $\sigma_1^2$ and $\sigma_2^2$ are treated as unknown parameters. Define the unknown parameter set as $\Theta  = \left\{ {\sigma _1^2,\sigma _2^2} \right\}$. The likelihood function of the received
pilot signal matrix ${{\bf{Y}}_{L \times N_p}}$ is given by
\begin{align}
&\mathcal{L}(\mathbf{Y}_{L \times N_p}; n_0, \Theta)\notag \\
&= \prod_{l = 1}^L \left[
    \prod_{n = 1}^{n_0} \frac{1}{\pi \sigma_1^2} \exp\left( -\frac{|y_l(n)|^2}{\sigma_1^2} \right)
\right. \notag \\
&\quad \left.
    \times \prod_{n = n_0 + 1}^{N_p} \frac{1}{\pi \sigma_2^2} \exp\left( -\frac{|y_l(n)|^2}{\sigma_2^2} \right)
\right]\notag \\
&={\left( {\frac{1}{{\pi \sigma _1^2}}} \right)^{{n_0}L}}{\left( {\frac{1}{{\pi \sigma _2^2}}} \right)^{({N_p} - {n_0})L}}\notag \\
&\times \exp \left( { - \frac{1}{{\sigma _1^2}}\sum\limits_{l = 1}^L {\sum\limits_{n = 1}^{{n_0}}  }{\left|{y_l}(n)\right|^2} - \frac{1}{{\sigma _2^2}}\sum\limits_{l = 1}^L {\sum\limits_{n = {n_0} + 1}^{{N_p}}  } {\left|{y_l}(n)\right|^2}} \right)\tag{7}
\label{eq7}.
\end{align}

By taking the natural logarithm of (\ref{eq7}) and omitting the constant terms unrelated to $n_0$ and $\Theta$, the log-likelihood function is given by
\begin{align}
\log \widetilde{\mathcal{L}}({\bf{Y}}_{L \times N_p}; n_0, \Theta) =& - n_0 L \log(\sigma_1^2)
    - \frac{1}{\sigma_1^2} \sum_{l=1}^{L} \sum_{n=1}^{n_0} \left| y_l(n)  \right|^2 \notag \\
&- (N_p - n_0) L \log(\sigma_2^2) \notag \\
    &- \frac{1}{\sigma_2^2} \sum_{l=1}^{L} \sum_{n=n_0+1}^{N_p} \left| y_l(n) \right|^2.
\tag{8}
\label{eq8}
\end{align}
To obtain the unknown parameters $\Theta$, we take the partial derivatives of the log-likelihood function with respect to $\sigma_1^2$ and $\sigma_2^2$, respectively,
\begin{equation}
\frac{{\partial \log \widetilde{\mathcal{L}}}}{{\partial \sigma _1^2}} =  - \frac{{L{n_0}}}{{\sigma _1^2}} + \frac{1}{{{{\left( {\sigma _1^2} \right)}^2}}}\sum\limits_{l = 1}^L {\sum\limits_{n = 1}^{{n_0}} {{{\left| {{y_l}(n)} \right|}^2}} }
\tag{9}
\label{eq9},
\end{equation}
\begin{equation}
\frac{{\partial \log \widetilde{\mathcal{L}}}}{{\partial \sigma _2^2}} =  - \frac{{L(N_p-n_0)}}{{\sigma _2^2}} + \frac{1}{{{{\left( {\sigma _2^2} \right)}^2}}}\sum\limits_{l = 1}^L {\sum\limits_{n = n_0+1}^{{N_p}} {{{\left| {{y_l}(n)} \right|}^2}} }
\tag{10}
\label{eq10}.
\end{equation}
By setting the above derivatives to zero, the ML estimates of $\sigma_1^2$ and $\sigma_2^2$ can be given by
\begin{equation}
\hat{\sigma}_1^2 = \frac{1}{L n_0} \sum_{l=1}^{L} \sum_{n=1}^{n_0} \left| {{y_l}(n)} \right|^2
\tag{11}
\label{eq11},
\end{equation}
\begin{equation}
\hat{\sigma}_2^2 = \frac{1}{L (N - n_0)} \sum_{l=1}^{L} \sum_{n = n_0 + 1}^{N} \left| {{y_l}(n)} \right|^2
\tag{12}
\label{eq12}.
\end{equation}
By substituting the estimated values $\hat{\sigma}_1^2$ and $\hat{\sigma}_2^2$ into (\ref{eq8}) and removing the constant terms unrelated to $n_0$, we can obtain
\begin{equation}
\log \widetilde{\mathcal{L}}({\bf{Y}}_{L \times N_p}; n_0) = - n_0 L \log(\hat \sigma_1^2)- (N_p - n_0) L \log(\hat \sigma_2^2).
\tag{13}
\label{eq13}
\end{equation}

Finally, the estimated values of $n_0$ is obtained by maximizing $\log \widetilde{\mathcal{L}}({\bf{Y}}_{L \times N_p}; n_0)$ over all possible points,
\begin{equation}
{{\hat n}_0} = \arg {\max _{{n_0} \in \{ 2,3, \ldots ,{N_p}-1\} }}\log \widetilde{\mathcal{L}}({{\bf{Y}}_{L \times {N_p}}};{n_0})
\tag{14}
\label{eq14}.
\end{equation}

Recalling the earlier assumption that $N_p>2\left |\tau\right |$, the value of STO $\tau$ can be determined based on the estimated $\hat n_0$ as follows. When $\hat n_0 < \frac{N_p}{2}$, this corresponds to early sampling, and the estimated STO is given by $\hat{\tau} = -\hat{n}_0$.  Otherwise, it corresponds to delayed sampling, with $\hat{\tau} = N_p - \hat{n}_0$. The STO estimation procedure is given in Algorithm \ref{table1}.

\floatname{algorithm}{Algorithm}
\renewcommand{\thealgorithm}{1}
\begin{algorithm}[H]
\setstretch{1}
\caption{MLE-Based STO Estimation}
\label{table1}
\renewcommand{\algorithmicrequire}{\textbf{Input:}}
\renewcommand{\algorithmicensure}{\textbf{Output:}}
\begin{algorithmic}[1]
    \REQUIRE \(L\), \(N_p\), received signal matrix \({\bf{Y}}_{L \times N_p}\)
    \ENSURE Estimated STO \(\hat{\tau}\)

    \FOR{\(n_0 = 2\) to \(N_p-1\)}
        \STATE Estimate \(\hat{\sigma}_1^2\) and \(\hat{\sigma}_2^2\) using (\ref{eq11}) and (\ref{eq12})
        \STATE Compute \(\log \widetilde{\mathcal{L}}({\bf{Y}}_{L \times N_p}; n_0)\) using (\ref{eq13})
    \ENDFOR

    \STATE Estimate the value of $\hat n_0$ using (\ref{eq14})
    \STATE Determine the estimated STO by
    \[
    \hat{\tau} =
    \begin{cases}
        -\hat{n}_0, & \text{if } \hat{n}_0 < \frac{N_p}{2} \\
        N_p - \hat{n}_0, & \text{otherwise}
    \end{cases}
    \]

    \RETURN \(\hat{\tau}\)
\end{algorithmic}
\end{algorithm}

Due to the limited number of pilot symbols used in Algorithm~\ref{table1}, the STO estimation error is also unavoidable and is defined as $\varepsilon = \tau - \hat{\tau}$, where $\tau$ denotes the true STO, $\hat \tau$ represents its estimated value.
\subsection{BER Performance}
With the estimated STO $\hat \tau$, STO compensation on the received signal is performed as follows
\begin{equation}
y^{\text{comp}}(n) = y^{\text{STO}}(n + \hat{\tau}),
\tag{15} \label{eq15}
\end{equation}
where $y^{\text{STO}}(n)$ denotes the received signal distorted by STO, $y^{\text{comp}}(n)$ represents the received signal after STO compensation.

To evaluate the performance improvement achieved by STO compensation, we adopt the ED detector as an example. The test statistic is defined as $\Gamma = \sum_{n=1}^{N} \left|y^{\text{comp}}(n)\right|^2$, and the detection threshold \footnote{In practice, the detection threshold can be determined by estimating the variance of the known bits without the need for CSI, as detailed in \cite{8007328}.}is given by\cite{8007328}
\begin{equation}
T^{\rm{ED}} = \frac{N P_0 P_1}{ P_0 +  P_1} \left[ 1 + \sqrt{1 + \frac{2( P_0 +  P_1)}{N( P_1 -  P_0)} \ln \frac{ P_1}{ P_0}} \right]. \tag{16} \label{eq16}
\end{equation}
Then the decision rule is expressed as
\begin{equation}
\left\{ {\begin{array}{*{20}{l}}
{{\Gamma}\mathop \gtrless \limits_{ \hat B(k) = 1}^{ \hat B(k) = 0}  {T^{{\rm{ED}}}},\ {\rm{if }}\ P_0 \ge P_1},\\
{{\Gamma}\mathop \gtrless \limits_{\hat B(k) = 0}^{\hat B(k) = 1} {T^{{\rm{ED}}}},\ {\rm{if }}\ P_1 \ge P_0}.
\end{array}} \right.
\tag{17} \label{eq17}
\end{equation}

\section{Simulation Results}
In this section, we present the numerical simulation results to validate the accuracy of the proposed MLE-based STO estimator and further assess its impact on BER performance improvement. The channel coefficients $h$, $\zeta$, and $g$ are generated according to a complex Gaussian distribution $\mathcal{CN}(0,1)$. The powers of the ambient RF source and noise are both set to 1. The mean absolute error (MAE) is calculated as the average of $\left|\varepsilon\right|$ over $10^5$ trials.

\begin{figure}
  \centering
  \includegraphics[width=0.44\textwidth]{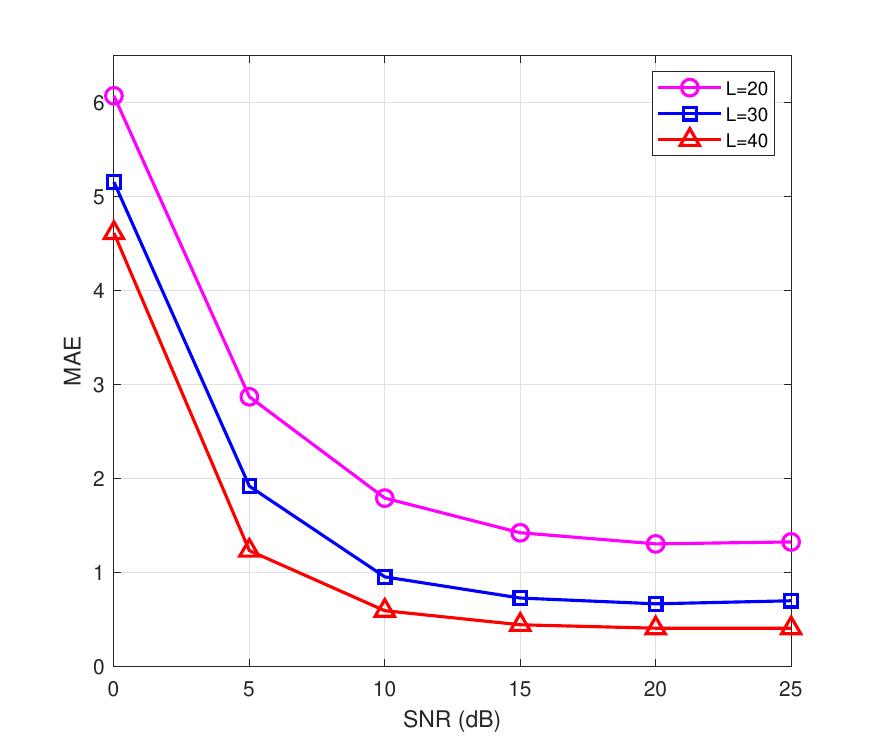}\\
  \caption{MAE versus SNR for $L=20,30,40$ for a fixed value $N_p = 30$. }\label{fig5}
\end{figure}

Fig. \ref{fig5} illustrates the relationship between the MAE and the SNR for different values of $L$, where $L$ denotes the number of pilot bit-pairs. The width of the sampling window is fixed at $N_p = 30$, and the STO is set to either $\tau = 10$ or $\tau = -10$. It can be seen that the MAE decreases significantly in the low to moderate SNR regime and gradually converges to an error floor at high SNRs. For example, at an SNR of $5\ \rm{dB}$, the MAE values for $L = 20,\ 30,\ 40$ are 2.8695, 1.9187, and 1.2333, respectively, and at an SNR of $15\ \rm{dB}$, the MAE values for $L = 20,\ 30,\ 40$ are 1.4230, 0.7285, and  0.4441. The error floor can be attributed to the limited number of pilot symbols. Additionally, the MAE decreases as $L$ increases, which indicates that a longer pilot sequence yields higher STO estimation accuracy.
\begin{figure}
  \centering
  \includegraphics[width=0.44\textwidth]{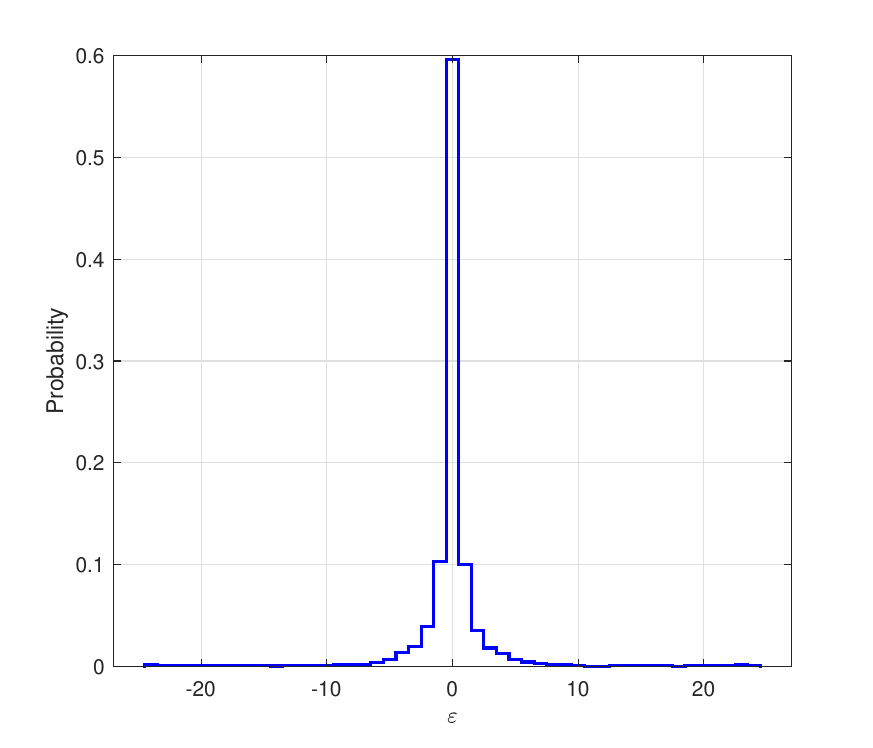}\\
  \caption{Empirical distribution of the STO estimation error at SNR $=\ 15\rm{dB}$, $L = 30$.}\label{fig6}
\end{figure}

Fig. \ref{fig6} presents the empirical distribution of the STO estimation error. The SNR is fixed at $15\ \rm{dB}$, both $L$ and $N_p$ are set to $30$, and $\tau$ is randomly selected from either the interval $[-10, -5]$ or $[5, 10]$. It can be seen that the estimation error is concentrated around zero, with a rapidly decaying probability on both sides. The result suggests that the proposed method achieves accurate and stable STO estimation for both $\tau < 0$ and $\tau > 0$.

\begin{figure}[t]
\centering
\subfigure[$N=50$]{
\label{fig7_a}
\includegraphics[width=0.23\textwidth]{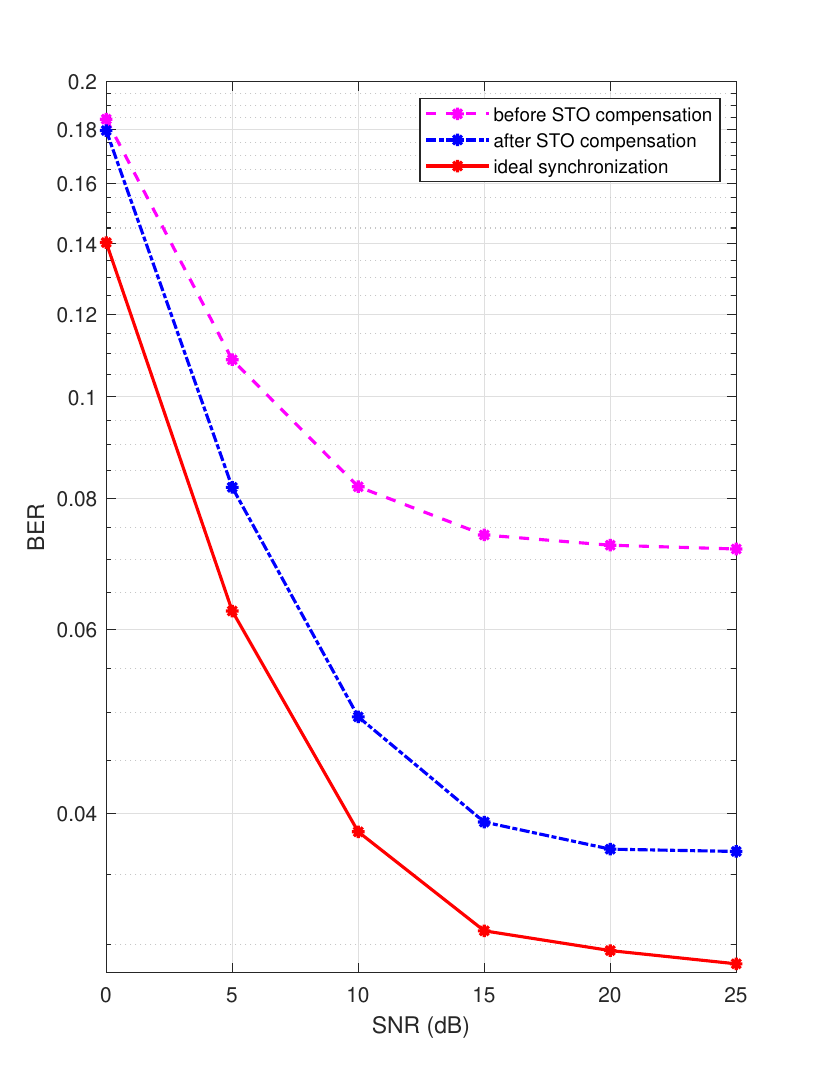}}
\subfigure[$N=100$]{
\label{fig7_b}
\includegraphics[width=0.23\textwidth]{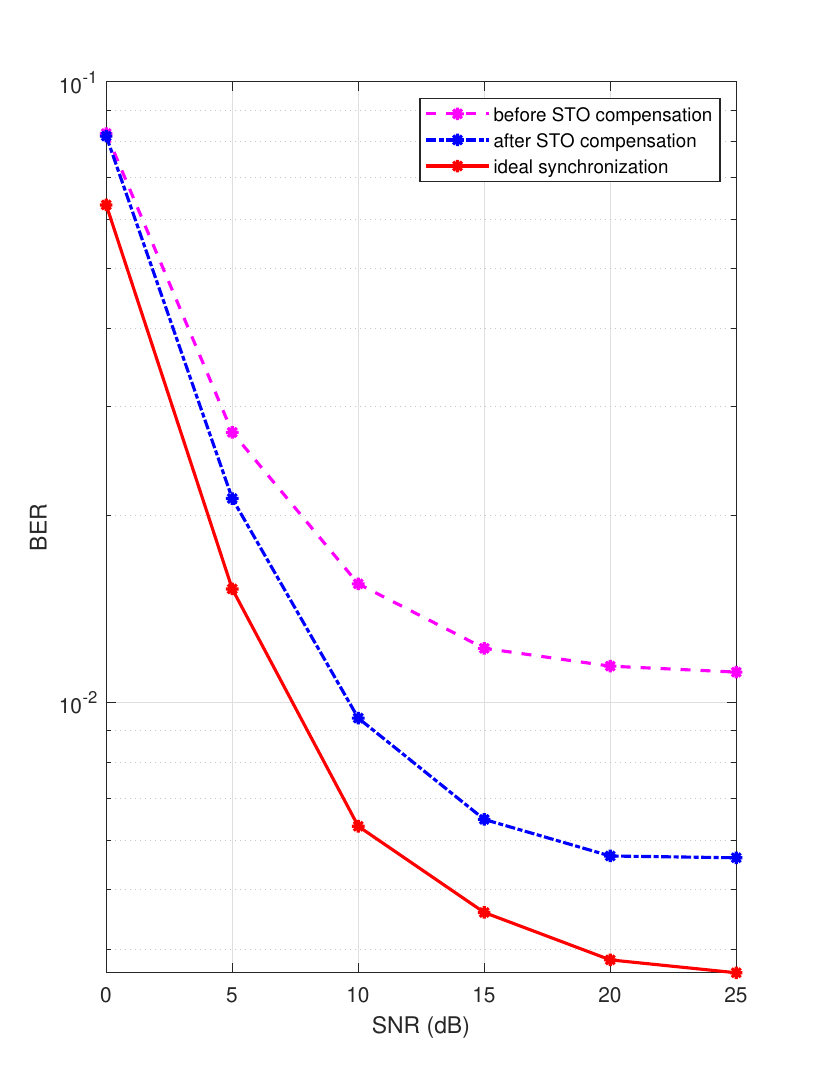}}
\caption{BER performance comparison under three conditions: before STO compensation, after STO compensation using the proposed method, and ideal synchronization for $N=50,100$.}
\label{fig7}
\end{figure}

Fig. \ref{fig7} compares the BER performance under three conditions: before STO compensation, after STO compensation using the proposed method, and ideal synchronization. The comparison results are shown for $N= 50$ and $N= 100$. The detection threshold is calculated based on (\ref{eq16}), and all other parameters are kept consistent with those in Fig. \ref{fig6}. It can be observed that the presence of STO significantly degrades the detection performance. This is mainly because STO may cause a deviation in the statistical characteristics of the received signal. After compensation with the estimated STO, the BER performance improves considerably, especially in the high SNR region, which demonstrates the effectiveness of the proposed STO estimation method. Even with STO compensation, a small performance gap remains compared to ideal synchronization. This is mainly due to the unavoidable estimation errors caused by the limited pilot length, which impair detection performance.
\section{Conclusion}
In this paper, we have investigated the non-ideal synchronization between the BD and BR in the AmBC system, where the ambient RF source is non-coordinated. Due to the presence of STO, the received signal corresponding to the current symbol contains partial samples from adjacent symbols, which may result in sampling errors that significantly degrade symbol detection performance. To address this issue, we have designed a specialized pilot sequence to introduce sampling errors and have proposed an STO estimation method based on MLE. Taking ED as an example, we have evaluated the BER performance after applying STO compensation based on the proposed estimation method. Simulation results have demonstrated that the proposed method can achieve accurate STO estimation and significantly mitigate the performance degradation caused by STO.


\bibliographystyle{IEEEtran}
\bibliography{refa}
\end{document}